\documentclass[]{aa}

\usepackage{txfonts}
\usepackage[normalem]{ulem}
\usepackage{graphicx}
\usepackage{natbib}
\bibpunct{(}{)}{;}{a}{}{,}
\usepackage{scrextend}		
\usepackage{tikz}
\usepackage{subcaption}
\usepackage{multirow}
\usepackage{hyperref}
\usepackage{amsmath}  
\usepackage{xfrac}   
\usepackage{amssymb}	
\usepackage{graphicx}
\usepackage{amsmath}  
\usepackage{xfrac}   
\usepackage{amssymb}	
\usepackage{dsfont}
\usepackage{txfonts}
\usepackage{mwe,tikz}\usepackage[percent]{overpic}
\usepackage{lipsum}
\usepackage{siunitx}
\usepackage{rotating}

\newcommand{\tC}{Type-C\xspace}

\begin{document} 

  \title{Are Low-Frequency Quasi-Periodic Oscillations seen in accretion flows the disk response to a jet instability?}

\titlerunning{Are LFQPOs seen in accretion flows the disk response to a jet instability?}

  \author{J. Ferreira \inst{1}
  \and G. Marcel \inst{2}
  \and P.-O. Petrucci \inst{1}
  \and J. Rodriguez \inst{3}
  \and J. Malzac \inst{4}
  \and R. Belmont \inst{3}
  \and M. Clavel \inst{1}
  \and G. Henri \inst{1}
  \and S. Corbel \inst{3}
  \and M. Coriat \inst{4}
	    }
         
   \institute{Univ. Grenoble Alpes, CNRS, IPAG, 38000 Grenoble, France
\and  Institute of Astronomy, University of Cambridge, Madingley Road, Cambridge, CB3 OHA, United Kingdom
\and AIM, CEA, CNRS, Universit\'e Paris-Saclay, Universit\'e de Paris, F-91191 Gif-sur-Yvette, France 
 \and   IRAP, Universit\'e de Toulouse, CNRS, UPS, CNES, Toulouse, France
             }

   \date{Received month dd, yyyy; accepted month dd, yyyy}

   \abstract
{Low Frequency Quasi-Periodic Oscillations or LF QPOs are ubiquitous in black hole X-ray binaries and provide strong constraints on the accretion-ejection processes. Although several models have been proposed so far, none has been proven to reproduce all observational constraints and no consensus has emerged yet.

We make the conjecture that disks in binaries are threaded by a large scale vertical magnetic field that splits it into two radial zones. In the inner Jet Emitting Disk (JED), a near equipartition field allows to drive powerful self-collimated jets, while beyond a transition radius, the disk magnetization is too low and a Standard Accretion Disk (SAD) is settled. In a series of papers, this hybrid JED-SAD disk configuration has been shown to successfully reproduce most multi-wavelength (radio and X-rays) observations, as well as the concurrence with the LFQPOs for the archetypal source GX 339-4.

We first analyze the main QPO scenarios provided in the literature: 1) a specific process occurring at the transition radius, 2) the accretion-ejection instability and 3) the solid-body Lense-Thirring disk precession. We recall their main assumptions and shed light on some severe theoretical issues that question the capability to reproduce LF QPOs. We then argue that none of these models could be operating under the JED-SAD physical conditions. 

We finally propose an alternative scenario where LF QPOs would be the disk response to an instability triggered in the jets, near a magnetic recollimation zone. Such a situation could account for most Type-C QPO phenomenology and is consistent with the global behavior of black hole binaries. The calculation of this non-destructive jet instability remains however to be done. If the existence of this instability is numerically confirmed, then it could also naturally account for the jet wobbling phenomenology seen in various accreting sources, around compact objets and young forming stars.}   

               
   \maketitle
%

\section{Introduction}

Black Hole X-ray Binaries (hereafter BH XrB) are binary systems where a stellar-mass black hole accretes matter from a low-mass companion star. The incoming mass forms an accretion disk around the black hole which is detected mainly in X-rays. All systems are X-ray transients, spending sometimes years in a barely detectable quiescence and suddenly increasing their luminosity by factors up to $10^6$ during months, until returning back to quiescence (see e.g. \citealt{remi06,done07,dunn10}). These episodes of intense activity are accompanied by drastic spectral changes, with two main canonical spectral states. In the hard state, the spectrum is dominated by a (cut-off) power-law like emission usually attributed to inverse Comptonization that peaks at a few 100 keV and shows a quite intense (10 to 20\% rms) rapid (subsecond) variability. In the soft state, the spectrum is characterized by a thermal black body emission of a few keV only and exhibits much less variability. A complete outbursting cycle can then be followed in a Hardness-Intensity Diagram (HID), where most BH sources make a q-shaped evolutionary track (see e.g. \citealt{dunn10}). Starting at low emission and in the hard state, an outburst consists first of a tremendous increase of luminosity in the hard state, a sharp transition to a softer (low hardness) state while maintaining a roughly constant luminosity, followed by a luminosity decrease while remaining in this soft state, another transition back to the hard state at roughly the same luminosity and a final decay back the initial state (finishing the q-shape). 

But the HID only probes accretion onto the black hole and does not convey all the phenomenology of these sources. It is now well known that jets, possibly in the form of bipolar self-confined outflows, are also emitted by the central regions. These jets are mainly detected in radio bands and show a clear correlation with the underlying X-ray emission, albeit only when the source is in the hard state (see e.g. \citealt{corb00,corb03,corb13a,gall03,merl03,falk04,fend04c} and the review by \citealt{fend14}). While explaining only the behavior seen in the HID was already quite challenging {\it per se}, including the jet production mechanism (and its associated power leakage in the hard states) remains an open issue. There have been several propositions of course (e.g. \citealt{esin97b,meye05,yuan05,ferr06a,bege14,kyla15}) but, to date, there is no clear consensus on which scenario is the most likely to explain the complex multi-wavelength (radio to X-rays) time dependent behavior of BH XrBs. 
 
The analyzis of the fast timing variations seen in hard X-rays offers however another independent means to grasp the BH XRBs phenomenology (see \citealt{mott16, ingr19} for recent reviews). The intrinsic variability of the sources is highly dependent of their spectral state, reaching 20\% in the hard states, going down to 1-10\% level in the hard-to-soft and soft-to-hard intermediate states and often consistent with zero in the soft state. Besides, the power density spectra have complex and state dependent shapes. On top of a broad band continuum, they can exhibit narrow features, called quasi-periodic oscillations or QPOs, of centroid frequency $\nu$ and extent $\Delta \nu$. There are three main types of low frequency ($\nu < 50$ Hz) QPOs that seem intimately linked to the spectral state \citep{remi99,case04,bell05}: Type-C, the most common, are detected in hard states, Type-B appear during (and define) the intermediate states while Type-A are rare, only seen in the high soft state. Type-A QPOs are very weak and broad with $Q=\nu/\Delta\nu < 3$ and a frequency $\nu_A\sim 6-8$ Hz. Type-B QPOs are more prominent and narrow, with $Q \geq 6$ and $\nu_B\sim 1-6$ Hz. Finally, Type-C QPOs are strong and narrow, with $Q\geq 10$ and a frequency $\nu_C$ ranging from mHz to $\sim10$ Hz. 

Although there is a time sequence from one QPO type to another, it is not clear whether all types should be explained by the same model. For instance Type-B QPOs are stronger in more face-on objects and reveal a time proximity with transient radio flares, which has lead to propose a link between their appearance and the launching of discrete relativistic ejections \citep{fend09,mott15,stev16,russ19,homa20}. But these two properties are not shared by all QPOs. Indeed, Type-A QPOs are detected in a jetless phase while Type-C QPOs are characterized by three extra properties \citep{remi06,mott16}. 
First, the frequency $\nu_C$ is independent of the energy band and source inclination, although always associated to the inner (hot) accretion flow (e.g \citealt{vEij17}). 
Second, the time evolution of the QPO frequency is tightly correlated with the variation of the soft ($<10$ keV)  source count rate and  hard X-ray power-law index, as well as the innermost radius of the cold (black body) accretion flow. However, the QPO frequencies are always much lower than the Keplerian frequency of the inner cold disk radius (e.g. \citealt{muno99,mark99,sobc00,rodr02,vign03,rodr04a,remi06,marc20}). 
Finally, phase lags are also commonly observed between different energy bands (e.g. 2-5 versus 13-30 keV, \citealt{case04}). The lag of a Type-C QPO is strongly correlated with its frequency and depends also on the object inclination: near zero but with an increasing hard lag (soft photons arriving first) with increasing QPO frequency for low inclination sources whereas high-inclination sources turn to soft lags at larger QPO frequencies (e.g. \citealt{mott15,vEij17,deRu19,zhan20a,ma21}).

Explaining Type-C LFQPOs is a challenging task and there have been already several models proposed in the literature. Yet, there is still no consensus as all QPO models are still facing major theoretical issues. Moreover, and more importantly, no model embraces the global picture of these accreting systems. Indeed, as intriguing as QPOs may be, they are only an epiphenomenon of the main accretion-ejection process. They must therefore be understood within, and be part of, a global framework that addresses also the other observational constraints namely, spectral evolution, jet formation and quenching.\\

Hereafter, we focus on such a framework, the hybrid disk configuration proposed by \citet{ferr06a}, which has been recently successfully confronted to observations in a series of papers \citep{marc18a,marc18b,marc19,marc20,marc22a,ursi20,mari20,barn22}. In this framework, a large scale vertical magnetic field $B_z$ is assumed to thread the inner disk regions, say below $10^4 r_g$ where $r_g=GM/c^2$ is the gravitational radius. At any given radius, its dynamical importance is measured at the disk midplane by the magnetization $\mu(r)= B_z^2/(\mu_oP_{tot})$, where $P_{tot}$ is the sum of the gas and radiation pressure. It has been argued that this magnetization is a decreasing function of the radius and that the disk is separated into two distinct regions \citep{ferr06a, petr08}. Beyond a transition radius $r_J$ the magnetization is low and the disk is assumed to be in a Standard Accretion Disk (hereafter SAD) mode, where most of the disk angular momentum is carried radially away by a MRI-driven turbulence \citep{shak73,balb91,balb03}. Although winds are possible and even expected when a large scale $B_z$ field is present, their influence on the disk energetics is rather small \citep{zhu18,jacq21}. As a consequence, the SAD region accretes at a highly subsonic speed and is optically thick. On the contrary, the region below $r_J$ and extending down to the innermost stable circular orbit $r_{ISCO}$ is in a Jet Emitting Disk (hereafter JED) mode, where a sizable fraction of the disk angular momentum is carried away vertically by two self-confined magnetically-driven jets. The JED model is a generalization and an extension of the \citet{blan82} jet model, addressing both the mass loading issue and causal connection with the underlying disk. This inner JED region is characterized by a magnetic field near equipartition (a constant $\mu$ around $0.1-0.8$), leading to a supersonic accretion speed \citep{ferr95,ferr97}. As a consequence, the JED region is much less dense than the outer SAD and becomes optically thin and geometrically slim. 
  
As estimated in \citet{marc18b}, the radial JED-SAD transition is achieved when the inner disk magnetization in the SAD reaches a value $\mu_c \sim 10^{-3}$. This critical value triggers the inward radial transition to a JED and leads to a sudden drop in density and increase in accretion speed as the jet torque becomes dominant. Postulated in 2006, this radial JED-SAD disk configuration seems to be qualitatively consistent with current 3D global MHD simulations, both in non-relativistic \citep{jacq21} and relativistic regimes (\citealt{igum03,tche11,McKi12,avar16,lisk20} to cite only a few)\footnote{In GRMHD simulations, this inner zone  has usually been termed a Magnetically Arrested Disk or MAD, a model initially defined as $B_zB_r^+/\mu_o \sim \rho \Omega_K^2 rh$, namely where the poloidal laminar magnetic field would be so strong that it would provide a support against gravity. Assuming $B_r^+\sim B_z$, this requires a disk magnetization $\mu \sim r/h$ (see \citealt{nara03} and references therein). On the other hand, a JED is a sub-Keplerian disk with $\mu$ of order unity, enough to launch centrifugally-driven jets. More numerical efforts must be done to assess the differences between a JED and a MAD.}.

This framework allows to reproduce most of the available data for the archetypical source GX 339-4. Indeed, by allowing  the disk accretion rate $\dot m$ and the transition radius $r_J$ to vary independently with time, 4 cycles of activity have been successfully reproduced (see \citealt{marc20} and references therein). It is remarkable that time evolutions of the pair $(r_J, \dot m)$ are able to reproduce each individual spectra and the HID of GX 339-4, but also the radio emission (although in a more qualitative way). Moreover, it has been also firmly established that the frequency of the detected \tC LFQPOs follows the scaling law 
\begin{equation}
\nu_{QPO} = \nu_K(r_J)/\chi 
\label{eq:rJ}
\end{equation} 
where $\nu_K(r_J)$ is the keplerian orbital frequency at $r_J$ and $\chi$ is a constant factor $\sim 100$ (actually varying from 70 to 130, depending on the outbursting cycle). This result, verified for more than two decades in frequencies, clearly advocates for a strong connexion between the physical source of the LFQPO and the transition radius $r_J$ \citep{marc20}.

To our knowledge, and despite several simplifications, this framework is currently the only one providing a clear picture addressing most of the observational accretion-ejection constraints. However, it is not clear yet if the observed timing properties, and in particular the LFQPOs, can fit inside this paradigm. This is the purpose of this paper. 

We first critically analyze in section 2 the main QPO scenarios provided in the literature: a specific process triggered at the transition radius, the accretion-ejection instability \citep{tagg99} and solid-body Lense-Thirring disk precession model \citep{ingr09}. We recall their main results but highlight also (sometimes severe) theoretical issues (see \cite{ingr19,marc21a} for observational issues). Putting these issues aside, it will be moreover argued that none of these scenarii can provide an explanation for Type-C QPOs fulfilling Eq.(\ref{eq:rJ}) within the JED-SAD framework. In section 3, we finally propose a novel scenario where the observed LFQPOs would be the inner JED response to a jet instability occurring away from the disk. It will be shown in Section 4 that such a situation not only provides a qualitative explanation for the LFQPO phenomenology, but it may also provide insights into the behavior of astrophysical jets. We conclude in section 5.

\section{Comments on some models for LF QPOs}
 
Before proposing a new scenario for the LFQPO (Sect.~3)  we first discuss in this section several existing models. We only address those which, in our view, are the most representative in the literature and refer instead the interested reader to the reviews of \citet{done07} and \citet{ingr19} for an exhaustive list. Note that all models require the existence of a transition radius usually associated to the innermost black body disk. When applicable, we discuss also some of these models in the context of the JED-SAD framework.

\subsection{A specific process occurring at the transition radius} 

Since the QPO frequency in hard X-ray emission is tightly correlated with the transition radius (labelled $r_J$ in the JED-SAD framework), it is natural to first look at a physical process that would be triggered at that specific location. The problem of course is the factor $\chi \sim 100$ which requires to look for a very slow, secular process. 

It is known for instance that any misalignment  between the black hole spin and the disk leads to a Lense-Thirring (hereafter LT) precession due to the dragging of inertial frames (see \citealt{bard75} and references therein). As a consequence, a test particle located at a radius $r$ from a black hole of dimensionless angular momentum $j$ would undergo a relativistic precession at the LT-frequency $\nu_{LT} (r)\simeq \frac{jc}{\pi r_g} (r/r_g)^{-3}$ which is steeply decreasing with the distance. Proposed initially for kHz QPOs in XrBs \citep{stel98,stel99}, one could also imagine that the outskirts of a hot inner flow, specifically a hot ring near the transition radius with the SAD, could provide an observable LFQPO. However, the ratio $\nu_{LT}/\nu_K(r_J)= 2j (r_J/r_g)^{-3/2} \propto  r_J^{-3/2}$ keeps decreasing with the distance. This is in contradiction with the observational constraint Eq.~(\ref{eq:rJ}), where this ratio remains roughly constant regardless of $r_J$. 

Motivated by the similarity between the observed power-density spectra of BH XrBs  and the response function to external perturbations of driven mechanical and electrical systems, \citet{psal00} computed the response of a narrow (of $\delta r/r\sim0.01$), geometrically thin accretion disk annulus to a broad spectrum of isothermal perturbations imposed outside of it. That ring corresponds to an abrupt change of disk properties, probably again the transition radius. They found that some resonances appear superimposed on the incoming spectrum, leading to observable QPOs. Since they neglected the radial pressure forces, the predicted mode frequencies depend mostly on the gravitational field around the compact object and only weakly on the hydrodynamic properties of the flow itself. As a result, the selected frequencies are related to the epicyclic and LT frequencies at the transition radius, which are both unable to explain LFQPOs due to the huge factor $\chi$.

\citet{kato00} proposed a model of trapped oscillations that would be triggered at the transition radius between the outer SAD and an inner Advection Dominated Accretion Flow or ADAF (\citealt{ichi77,nara94}, see also the review by \citealt{yuan14}). Such a situation implies a very narrow transition zone where the rotation profile needs to become slightly super Keplerian \citep{abra98}. As a consequence the epicyclic frequency $\kappa^2$ becomes negative, leading to a local instability (Rayleigh criterion, when the stabilizing effects due to the strong inhomogeneities are negligible). The authors then argue that the slowly growing amplitude of the perturbations would remain trapped around the transition radius, leading to local QPOs.

Assuming a particular radial profile for the disk temperature within the transition region, they solved the local dispersion relation of these inertial-acoustic modes. They showed that it is possible to fine-tune the radial profile so that the eigenvalue could match the QPO frequency at a given transition radius. However, and this is quite uncomfortable, this fine-tuning must be done for every radius and for a given set of disk parameters. It seems therefore difficult to reconcile this result with the generic behavior deduced from observations\footnote{This fine-tuning has nothing to do with a "p-disk model", which is a multi-blackbody disk model where the temperature exponent is left free \citep{mine94,kubo05}.}. Not only there is no reason why such a fine-tuning should occur, but the model relies also on strong approximations. Of particular importance is the neglect of the accretion motion, which would unavoidably lower the local growth of perturbations by advecting them.

This remark is even more critical within the JED-SAD framework, which assumes the presence of a large scale $B_z$ field everywhere. Whether or not  such an instability would be present near $r_J$ in this case requires a novel investigation. However, the disk material undergoes a transsonic transition at $r_J$ \citep{ferr06a,marc18b,scep19}, since the SAD accretes at subsonic speeds whereas the JED is supersonic. Moreover, in order for a JED to maintain a constant disk magnetization $\mu$, any extra magnetic flux that would be carried in by the accreting flow must be expelled out. This is exactly what happens in some GRMHD simulations, where a magnetic Rayleigh-Taylor instability is seen to grow up and expel the magnetic flux excess as magnetic bubbles \citep{McKi12,marsh18}. Quite interestingly, this is done quasi-periodically but on a time scale comparable to a few Keplerian orbits at $r_J$, inconsistent with LFQPOs. The existence of such a messy, transsonic JED-SAD transition casts therefore serious doubts on the possibility to allow the development of any secular instability at that location.

\subsection{The accretion-ejection instability}

The accretion-ejection instability (AEI) has been first studied by \citet{tagg99} in the context of accretion disks threaded by a large scale $B_z$ field, and then proposed to explain QPOs by \citet{varn02a} and \citet{rodr02}. A non-axisymmetric ideal MHD instability is found to be triggered whenever the vertical field reaches equipartition along with some radial profile. These spiral waves travel back and forth between the disk inner radius $r_{in}$ (initially assumed to be the innermost SAD radius) and some radius $r_{co}$, defined as the co-rotation between the wave phase speed and the disk material rotation. The waves transport angular momentum outwardly, allowing the "cavity"  between $r_{in}$ and $r_{co}$ to accrete rapidly. Although spiral waves are also emitted beyond $r_{co}$, accretion in this outer region is nevertheless assumed to carry on thanks to the usual MRI-driven turbulence. At the co-rotation radius, waves are evanescent but a resonance with a vertical Alfv\'en wave allows the transfer of some energy and angular momentum in the vertical direction. A magnetized azimuthally localized vortex is thus expected to grow at $r_{co}$, opening the possibility for some ejection there (but not demonstrated yet). Such a vortex is invoked to be the locus of enhanced dissipation (a hot spot), leading to a QPO with a frequency $\nu_K(r_{co})$, as long as some geometrical occultation is also present of course {(so QPOs can be seen only in high inclination sources)}. 

At first sight, the AEI would perfectly fit at the interface between the JED and SAD zones, due to the build-up of a large scale $B_z$ field near equipartition in the JED. But in order for the AEI to reproduce the LFQPO behavior encapsulated in Eq.~(\ref{eq:rJ}), the ratio of the co-rotation radius to the innermost disk radius (equal to $r_J$ within the JED-SAD framework) must then be $\chi^{2/3}>>1$ (independent of $r_J$). Such a distance between the inner boundary and the co-rotation radius cannot be achieved within the AEI framework (see for instance Fig.~4 in \citealt{varn02a}). Indeed, as the magnetization increases, the wavelength becomes larger, widening the "cavity" and 
pushing out to larger radii the co-rotation radius. But this leads also to a widening of the forbidden zone around the corotation radius and the amplification mechanism becomes less efficient. As a result, the growth rate has a maximum for $\mu$ only close to 1 \citep{tagg99}. In other words, the factor $\chi \sim 100$ for Type-C QPOs cannot be achieved within the AEI framework for reasonable values of the magnetic field.    
   
It must also be realized that the complex situation invoked for the AEI has been computed with a highly simplified setup so far. For instance, the toroidal component of the magnetic field has been neglected, which forbids of course the launching of jets and winds that would seriously affect the disk angular momentum transport. The unavoidable MRI-driven turbulence has been also discarded, while it provides both another torque and wave dissipation, two ingredients that would again lower the efficiency of the mechanism. On the other hand, despite the assumption of a perfect wave reflection at the inner disk boundary, the maximum AEI growth rate is already quite low, namely of order $V_A/r \sim \mu^{1/2} \Omega_K h/r$ only \citep{tagg99}. This means that the physical ingredients listed above (jets, turbulence) would probably seriously lower it or even quench the instability. 
Finally, the probable presence of a magnetic Rayleigh-Taylor instability at the edge of the JED, as seen in MAD simulations \citep{McKi12,marsh18} questions the very existence of the AEI at that location. Because of all these extra ingredients, assuming a perfectly reflecting boundary at $r_J$ appears rather dubious indeed.

Given all these difficulties and in particular the unreachable factor $\chi$, we think that AEI is very unlikely the source of Type-C QPOs within the JED-SAD framework. However, for Type-A and/or Type-B LFQPOs, that are observed only when $r_J$ is near the ISCO and relativistic effects important (see \citealt{varn12} and references therein), the AEI may remain a possible candidate.

\subsection{Solid-body Lense-Thirring disk precession model} 

The solid-body precession of the inner disk due to the Lense-Thirring effect is certainly the most popular model invoked to explain Type-C LFQPOs \citep{ingr09, ingr19}. This model relies on a geometrical general relativity effect, occurring whenever the black hole spin is misaligned with the disk axis. The resulting frame drag causes important structural changes in the surrounding accretion disk (precession) as it struggles to adapt its equatorial plane to the periodic changes in the local gravitational field. The model of \citet{ingr09} for QPOs has been designed within the framework of \citet{esin97b}: an inner hot, optically thin and geometrically thick flow (ADAF) is settled until a transition radius $r_t$ beyond which the disk becomes the usual cold, optically thick and geometrically thin SAD. Within this framework, a LFQPO correlated with $r_t$ (hence also with the disk black body) could arise if the inner geometrically thick disk (ADAF) modulates the X-ray flux. But this can only happen if a significant portion of the ADAF's volume is precessing as a {\it solid body}, namely with a unique precession frequency $\nu_{prec}$. 

As shown by \citealt{ingr09}, if that volume extends down to the innermost stable orbit, not only the precession frequency $\nu_{prec}$ would exhibit a too strong dependence on the black hole spin, but it would also provide frequencies far too high. As a consequence, it is usually assumed that the innermost disk regions, up to a few $r_g$, are actually aligned with the black hole spin, probably due to a Bardeen-Petterson effect \citep{bard75,papa95,lubo02}. Alignment of the inner zone is actually always seen in all numerical simulations of tilted black holes (see e.g. \citealt{frag07,frag09c}), although the presence of large scale magnetic fields seems to seriously modify the estimate of this inner radius and even question the physical mechanism \citep{mcKi13,krol15,lisk18,lisk19,chat20,lisk21}. Nevertheless, assuming an inner radius of a few $r_g$ and a solid-body LT-precession up to an arbitrary transition radius $r_t$, \citet{ingr09} have shown that the precession frequency $\nu_{prec}(r_t)$ could indeed cover the observed frequency range of Type-C QPOs from $\sim0.01$ to $\sim 10$ Hz. Note that the actual value of $\nu_{prec}$ is a direct function of the external radius $r_t$, which remained a free parameter in this work. Several other observational features related to QPOs have been obtained within this framework (see the review of \citealt{ingr19}).

However, as appealing as this scenario is, the proof that a solid-body LT-precession is settled over a significant radial range remains controversial in numerical GR simulations. Indeed, while solid body precession has been clearly shown in hydrodynamical flows (see eg. \citealt{frag05,dyda20}), the trend is much less clear when magnetized flows are considered. There has been some works claiming that there is indeed a volume undergoing a solid-body LT-precession (e.g. \citealt{frag07,lisk18,lisk19}) and others where no evidence has been found (eg. \citealt{mcKi13,sora13,krol15,chat20}). The main reason of this discrepancy is probably the difference in the large scale magnetic field accumulated in the central disk regions (namely the value of the achieved disk magnetization $\mu$). Indeed, not only a vertical field triggers an MRI-driven turbulence but it also launches jets from the disk. And it is well known that both effects, turbulence and jets, produce extra torques that may prevent the enforcement of solid-body precession by the Lense-Thirring torque \citep{sora13,mcKi13}. It is also possible that high tilts ($65^o$) would be able to tear off the inner disk up at a radius $r_t$ and enforce a solid-body precession \citep{lisk21}. However, it would be a rather strong implication of the model if, in order to reproduce LFQPOs, high tilts would always be required.

In any case, we would like  to stress that providing the proof that a significant portion of the disk, say from $\sim 5 r_g$ to $r_{out}\sim 20$ or even up to $100 r_g$, is actually precessing as a solid-body is a formidable numerical task. It would require for instance to show that the local precession angle is the same for the whole range of radii, and that it evolves linearly in time such that $P(t)- P(0)= \dot Pt$, with $\dot P=2\pi \nu_{QPO}$ allowing to measure the QPO frequency (see e.g. Fig~2 in \citealt{dyda20}). But this demands to maintain disk conditions at $r_{out}$ (i.e. accretion rate) rather constant for a duration $\Delta T$ that must be at the very least 3 or 4 times the QPO period, namely $\Delta T\sim 1-3\, (r_{out}/r_g)^{3/2} 10^3 r_g/c$. For instance, a 0.1 Hz QPO would require a converged simulation that would last for a few times $10^5 r_g/c$ for a 10 solar masses BH. While such long times have been recently achieved in a few simulations (e.g. \citealt{chat20,lisk21}), constant disk conditions in the outer regions are still not met (to our knowledge). This situation is due to the initial conditions, which do not include a steady outer accretion disk. It is therefore very difficult to assess whether a solid-body LT-precession is actually settled in and clearly deserves further numerical simulations, long enough and tailored to maintain an outer cold thin disk.

Notwithstanding theses current difficulties, one could ask if this scenario could fit within our JED-SAD framework. It requires of course the black hole spin to be misaligned with the disk angular momentum vector, and there is no reason why it should not be (although large tilts are not expected to always be the rule). As a result of the LT-torque, the innermost  disk region (up to a few $r_g$ according to GRMHD simulations) should always be aligned with the BH spin, whether the disk is in a JED or in a SAD accretion mode (note however that this alignment may be modified from the usual Bardeen-Petterson mechanism by the presence of large scale magnetic fields \citealt{lisk19,lisk21}). But what happens then beyond this innermost aligned region, when there is an inner JED settled up to a large $r_J$?  

According to \citet{mcKi13} and references therein, one can answer this question by comparing the LT-torque with the other local torques acting on the disk. The local LT torque acting on a disk annulus can be estimated as $\Gamma_{LT} = \Omega_{LT} L \sin\beta$, where $\Omega_{LT}$ is the LT-precession pulsation, $L=  \Sigma \Omega_K r^2$ is the disk angular momentum per unit area and $\beta$ the black hole tilt angle (see e.g. \citealt{mcKi13} and references therein). In a JED, the dominant torque is the magnetic braking provided by the two jets and writes $\Gamma_{jets}= -2r \frac{B_\phi^+B_z}{\mu_o}$, which leads to a ratio
\begin{equation}
\frac{\Gamma_{LT}}{\Gamma_{jets}} =  \sin\beta \frac{2j}{q\mu} \frac{r}{h}  \left (\frac{r}{r_g}\right)^{-3/2}
\label{eq:torque}
\end{equation}
In this expression, $q= -B_\phi^+/B_z$ is the magnetic shear measured at the disk surface, $\mu$ is the disk magnetization measured at the disk midplane (note that it does not include turbulent fields) and $h/r$ the local disk aspect ratio. Since a hot JED verifies $q\mu \sim 1$ and $h/r\sim 0.1$ \citep{marc18a}, this expression shows that the LT torque should never be dominant in a JED beyond a few $r_g$. Please note that this conclusion stems only from our assumption of the existence of a near equipartition laminar magnetic field. 

This result can also be understood in other terms. In order for a solid-body LT-precession to take place, the disk must behave as a whole entity. This situation requires that a strong causal connection is maintained between its two radial boundaries. Namely, that bending waves can propagate back and forth and enforce the same precession rate over the entire volume. This "wave-like" regime has been estimated in hydrodynamical situations to require $\alpha_v < h/r$, namely a turbulent Shakura-Sunyaev $\alpha_v$ parameter smaller than the disk aspect ratio \citep{papa95,lubo02}. Since MRI provides a scaling $\alpha_v \sim 10 \mu^{1/2}$ \citep{salv16}, it shows that weakly magnetized ($\mu < 10^{-4}$), hot ($h/r>0.2$) flows could be in this regime and undergo indeed a solid body LT-precession. This was the situation envisioned initially (with an ADAF as the hot inner flow) and possibly achieved in some weakly magnetized GRMHD numerical simulations. But a JED has a near equipartition field ($\mu \sim 1$) and accretes at supersonic speeds due to its dominant jet torque. This challenges any upstream wave propagation and certainly forbids thereby the establishment of a solid body LT-precession over its entire volume
\footnote{Note that the JED thermal balance calculations done by \citet{marc18b} lead to a disk aspect ratio $h/r$ varying slowing with the radius (consistent with both the ion and electronic temperature profiles). On the other hand and for simplicity, a constant accretion Mach number (larger than unity) has been assumed in the JED, consistent with self-similar calculations. In a more realistic situation one would expect also a varying accretion Mach number. However, observations do require a supersonic accretion flow throughout the hot accretion flow \citep{marc21a,kawa22}, which means that the accretion speed (hence the jet torque) must also adapt. Thus, although there is a caveat of our calculations here, we believe it does not affect the main arguments presented in this section.}. 

How much of the JED volume, located beyond the innermost region aligned with the BH, could be actually undergoing a solid-body precession remains an open issue. We stress however that in order to recover the observational correlation encapsulated in Eq.(\ref{eq:rJ}), that volume should reach a radius $r_t \equiv r_J$, which is doubtful according to both the previous causal argument and torque estimates. In the case of high tilts leading to disk tearing at a radius $r_t$ \citep{lisk21}, this radius should then also coincide with the transition to an outer SAD accretion mode. Within our JED-SAD framework, it is unclear to us why $r_J$, which marks the transition from $\mu$ near unity in the JED to $\mu \sim 10^{-3}$ in the inner SAD, should always be coincident with the tearing radius $r_t$. 

Note finally that \citet{marc21a} reached a similar conclusion based only on observational constraints. Indeed, in order to reproduce the observed X-ray spectra during the most luminous hard states, the hot flow must accrete at sonic to supersonic speeds (see e.g. \citealt{marc21a,kawa22}), unreachable with typical viscous torques. Since Type-C QPOs are prevalent in these luminous states, they concluded that solid-body LT-precession is unlikely the driving mechanism. 
 
The existence of a BH-aligned precessing inner JED region translates of course into an inner jet precession and could therefore contribute to produce Type-A and/or Type-B QPOs \citep{stev16,lisk19,lisk21,kyla20,ma21}. And it could also provide the geometrical effect that is invoked to explain the influence of the source inclination on both the QPO amplitude and lags (e.g. \citealt{mott15,heil15,vEij17}). But according  to the above discussion, we doubt that solid body LT-precession could be the generic physical mechanism responsible for Type-C QPOs in all XrBs and believe that another mechanism needs therefore to be found (see also \citealt{nath22}).

\section{LFQPOs as the disk signature of a jet instability}

Type-C LFQPOs are detected in the hard energy band which is associated to a corona or hot inner flow (see however discussion in \citealt{rodr04b,rodr08}), but their frequencies show a tight correlation with the inner standard accretion disk radius, which is the JED-SAD transition radius $r_J$ in our view (Eq.\ref{eq:rJ}). The difficulty is to reconcile this correlation with a factor $\chi \sim 100$, which requires a secular process.
 
In the JED-SAD framework, two bipolar self-confined jets are magnetically launched from the inner JED. Thus, instead of looking for a secular instability within the disk itself, we propose that these QPOs are the disk response to some instability triggered in the jets themselves, away from the disk. This idea could naturally reconcile the Type-C QPOs' low frequency (long-term or large scales behavior) with their apparent link with the transition radius $r_J$. There are several aspects that must be considered: (i) How can a jet instability still impact the underlying disk ? (ii) Why would it have an influence on the JED spectrum ? (iii) What kind of jet instability would then be necessary ? 

\subsection{Causal connection with the underlying disk}

Let us first assume that jets  launched below $r_J$ are indeed prone to some  global instability. Since jets are clearly observed up to large scales, this instability must not lead to jet disruption. We thus only require that its non-linear stage leads to some local plasma and electric current re-organisation, ending up mostly into jet wobbling. It is that wobbling that defines a frequency, which is then expected to be conveyed backwards to the disk through waves. If this instability is triggered in the causally connected jet region, namely before the Fast-Magnetosonic (FM) surface, then FM waves can indeed propagate upstream and reach the disk (on a time of the order of the Keplerian orbital time scale at $r_J$, e.g. \citealt{ferr04}) so that we expect them to lead only to some broad band noise. But much longer time and/or spatial scales must be at play in order to trigger Type-C QPOs. 

\begin{figure}
   \centering
   \includegraphics[width=\columnwidth]{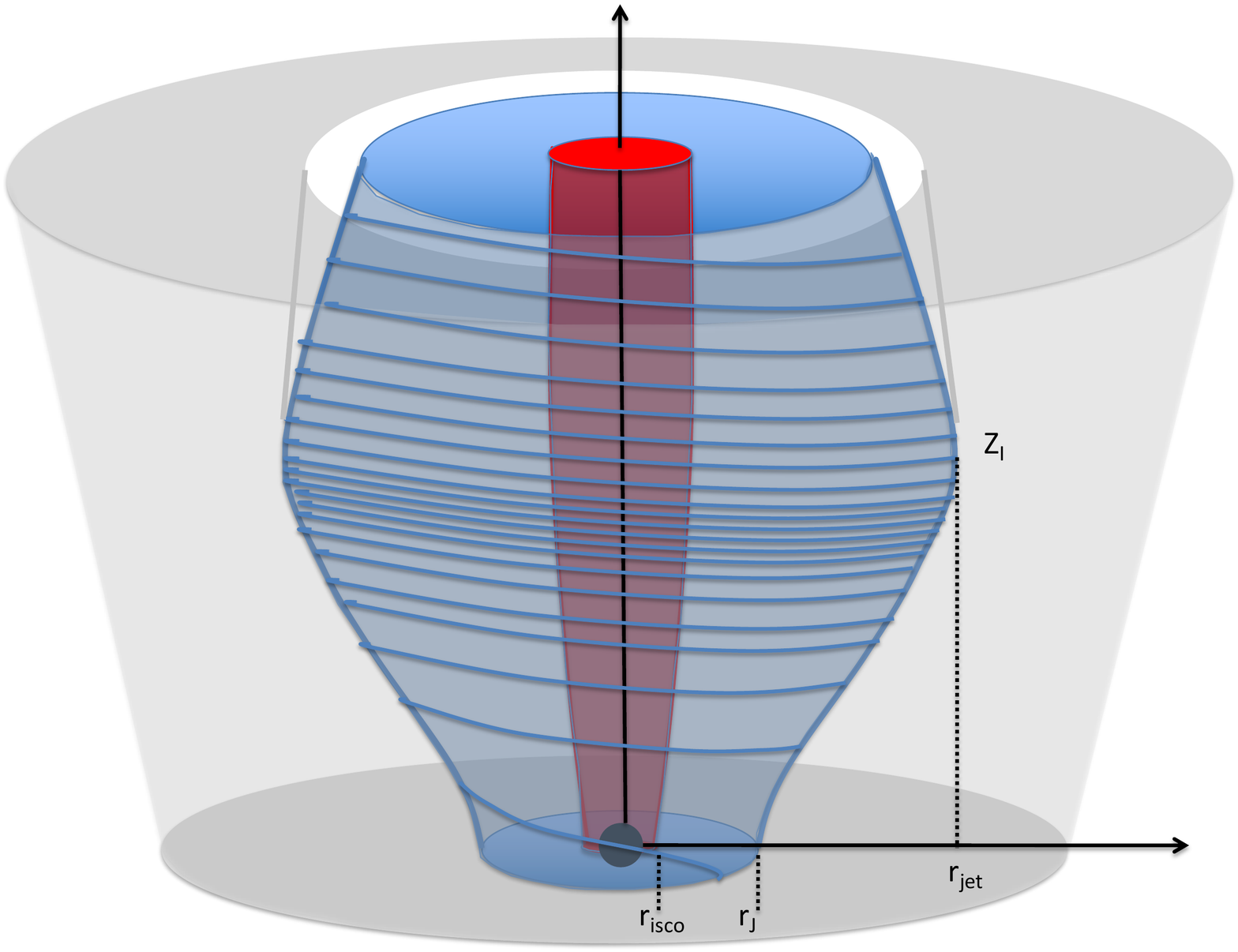}
      \caption{Sketch of the JED-SAD hybrid disk configuration and its associated outflows in the hard state. The Jet Emitting Disk is settled from the ISCO $r_{isco}$ up to the transition radius $r_J$, beyond which a Standard Accretion Disk is established. A vertical large scale magnetic field is threading the whole region, up to the axis. The red zone is the relativistic Blandford-Znajek spine, the blue zone is the sub to mildly relativistic Blandford-Payne jet launched from the JED and the grey shaded area is the outer magnetic sheath, with only unsteady sub-Alfv\'enic outflow (wind). The optimal location for the jet instability is the altitude $z_I$ where the outer magnetic surface anchored at $r_J$ is starting to recollimate toward the axis, defining the jet radius $r_{jet}$.}
         \label{fig1}
   \end{figure}

We are thus lead to assume that this jet wobbling occurs beyond the FM surface. In that case, waves can no longer propagate upstream {\em within the jet}. However, the JED-SAD framework requires the existence of a large scale vertical field threading the whole accretion disk. Although the super-FM jet is defined with the magnetic flux threading the JED, there is still magnetic field around it, threading the SAD. Such a field defines a magnetic sheath inside which the inner jet is propagating (see sketch in Fig.\ref{fig1}). This outer sheath being sub-FM (as for instance in the simulations of \citealt{murp10}), waves can still propagate downwards and reach thereby the disk. The path followed by these waves is not straightforward as the medium is inhomogeneous and waves are known to undergo some refraction. However, we expect modes to propagate preferentially along the sheath  down to the transition radius $r_J$ (much alike a surface mode). Moreover, once these waves reach the sub-FM zone of the inner jet, they will start to act as a lateral source localized at the limiting surface anchored at $r_J$. This will lead, in turn, to perturbations able this time to propagate within the jet and reaching the whole JED extension. 

Our main assumption is therefore the existence of a jet instability triggered at a distance $z_I$ leading to a subsequent jet wobbling. This lateral jet displacement bounces back on the magnetic sheath, which triggers the propagation of perturbations. How long it takes for these perturbations to reach the disk is a difficult question, as it requires to follow the path of these waves. Note however that once the instability is triggered, the perturbations in the jet itself are advected downstream, leaving behind the same physical conditions that have led to the triggering of the instability. As a consequence, it will grow again near $z_I$ and lead to another unstable regime with a global jet displacement. As long as the conditions for the jet instability are met, one should thus see this going on, defining thereby a jet wobbling frequency $\nu_I$ which is mostly related to the time scale required for the lateral jet displacement. It is that frequency that will be observed as a LF QPO in the disk.

\subsection{Impact on the accretion flow: LF QPOs}

The jet instability acts like a hammer hitting the sheath with a characteristic frequency $\nu_I$. The information (wobbling frequency and released energy) is transported backwards by FM waves travelling along the magnetic sheath surrounding the jet near $r_J$. A fraction of this energy is then expected to be spread all over the JED, through a lateral "shower" of FM modes triggered at the magnetic sheath near the disk. How is that supposed to affect the emitted hard X-ray spectrum and lead to an observable LFQPO? 

The fluctuations, typically 1-20\% seen in hard X-rays, can be understood through two independent processes, both related to these incoming waves. Since they are FM waves, some vertical compression of the hot JED material is naturally expected, leading to some enhanced dissipation (thus emission) at the disk surface. The second aspect is related to the possible modifications of the jet torque acting upon the disk. These waves are indeed expected to introduce also fluctuations on the local toroidal magnetic field, hence on the torque due to the jet.  As a consequence, one should see fluctuations of the JED accretion rate and on the related release of accretion energy, responding with the same quasi-periodic frequency $\nu_I$. 

However, to get a proper QPO with a rather large quality factor, one needs to have an almost resonant cavity. As shown by \citet{caba10}, the JED itself may play this role, acting also as a low bandpass medium: its response to a white noise excitation is a flat-top noise power spectral density at low frequencies and a red noise at high frequency. In other words, if the jet acts like a hammer, the bell would be the JED\footnote{This is somehow similar to the \citet{psal00} approach: the hot inner JED acts as a filtering cavity of all incoming perturbations.}. A QPO is therefore expected to arise at a frequency $\nu_I$, the same at each energy bin and related to the dynamical (epicyclic or keplerian) frequency at $r_J$ (see below). It is however unclear how the existence of the outer magnetic sheath and intrinsic fluctuations of the dissipation within the JED will affect the calculations done by \citet{caba10}. This should deserve further investigations.

\subsection{Which instability could lead to jet wobbling ?}

An MHD jet is a helical magnetic structure that can be seen as an ensemble of poloidal magnetic surfaces, nested around each other and in pressure equilibrium with the surrounding medium. In our JED-SAD framework, the jet is mostly the super-FM outflow emitted by the underlying JED established between the ISCO $r_{isco}$ and $r_J$ (blue region in Fig.\ref{fig1}). Below $r_{isco}$, magnetic field lines are brought in and concentrated over the plunging region and black hole horizon (red region in Fig.\ref{fig1}), leading to the production of a relativistic tenuous spine through the Blandford-Znajek process \citep{blan77}. It is however unclear whether the existence of this spine has any dynamical impact on the outer Blandford-Payne jet. Indeed, radio observations can be accounted for using only the power carried by this outer flow component \citep{marc18b,marc19,marc20}. But, in any case, if some jet instability is to give rise to a LFQPO related to $r_J$, it must involve the distant thus non-relativistic jet region. We thus neglect below the effect of the spine (see however \citealt{barn22}).  

Jets are prone to many instabilities such as pressure driven (PD), current driven (CD) and Kelvin-Helmholtz (KH) instabilities (plus any combination of them, \citealt{appl92,baty02}, for a review see e.g. \citealt{hard13}). All these instabilities can either trigger radially localized, internal and surface modes or long-wavelength body modes. The final non-linear outcome of these instabilities can go from a simple internal redistribution of jet quantities (leading to a non-linear jet stabilization) to the production of internal shocks and/or MHD turbulence. In this latter case, a significant energy dissipation may eventually end up to a complete disruption of the jet on a finite distance. There is no consensus on which instability would be dominant in a jet as their existence and relative strength (growth rate) depend on the jet radial stratification, which is still poorly understood. 

In this paper, we envision a non-destructive instability that gives birth to a long-wavelength instability at an altitude $z_I$, namely a global body-mode that involves lateral displacements of the whole outflow of radius $r_{jet}$. The jet radius $r_{jet}$ is the radius achieved at $z_I$ of the magnetic surface anchored at $r_J$. As discussed above, this displacement will bounce at the jet/magnetic sheath interface and trigger upstreaming (toward the disk) propagations of MHD disturbances. The jet displacement itself is advected downstream the flow and become potentially detectable farther out as jet wobbling. This allows another sequence (instability, jet displacement, bouncing) to set in at  $z_I$, so that the whole process will be quasi-periodic with a characteristic frequency $\nu_I$. The altitude $z_I$ is therefore the lowest distance from the source where such an instability can take place. The frequency $\nu_I$ is related to the time scale involved for allowing FM waves to travel across the jet radius at $z_I$, so that it can behave as a whole (body mode). The frequency can thus be estimated as 
\begin{equation}
\nu_I \sim \left . \frac{V_{FM}}{2r} \right |_{r_{jet}}
\label{eq:nu_i}
\end{equation}
where $V_{FM}$ is the FM phase speed at the jet radius $r_{jet}$. Assuming that the jet is cold and dominated by the toroidal magnetic field leads to 
\begin{equation}
V_{FM}\simeq V_{A\phi} = \frac{B_\phi}{\sqrt{\mu_o\rho}}=  \frac{m}{m^2-1}\left (1- \frac{r_A^2}{r^2} \right ) \Omega_* r \simeq \frac{\Omega_* r}{m}
\end{equation}
evaluated at $r=r_{jet} \gg r_J$ \citep{pell92}. In this expression, $\Omega_*\simeq \Omega_K(r_J)$ is the angular velocity of the magnetic surface (MHD invariant), $r_A$ is the Alfv\'en radius where the flow becomes super-Alfv\'enic and $m= u_p/V_{Ap}$ is the Alfv\'enic Mach number, assumed to be much larger than unity. 

In order for the characteristic frequency $\nu_I$ to be identical to the LFQPO frequency $\nu_C$, the condition  
\begin{equation}
m \sim  \pi \chi
\label{eq:m}
\end{equation}
on the Alfv\'enic Mach number $m$ must be fulfilled, where we made use of Eq.(\ref{eq:rJ}). This is possible as long as magnetically driven jets from accretion disks are able to reach a value $m$ between $\sim 220$ and 400, for $\chi$ ranging from $\sim 70$ to 130 \citep{marc20}. This is indeed the case, as shown for instance in Fig.~8 in \citet{ferr97}. It corresponds to tenuous MHD solutions that become super-FM (namely $n= u_p/V_{FM}$ of a few) and that expand the most, with $r_{jet}/r_J$ reaching values of a few hundreds up to a thousand (see his Fig.6). Note that Eq.(\ref{eq:m}) is not a fine tuning but instead a selection (and a property) of the underlying MHD solution established within the JED. The fact that such MHD solutions have been already computed is quite promising for our scenario.   

Jets are highly inhomogeneous media, with both tremendous magnetic and velocity gradients whose effects on the development of an instability are not fully understood yet (see e.g. \citealt{bond87,birk91a,kers00,fran00,appl96,mizu09,mizu11,mizu14,kim15,kim16,kim18}). We note however that no instability is expected below the Alfv\'en surface. Indeed, the stabilizing magnetic tension due to the poloidal magnetic field is be too important. Besides, just after the Alfv\'en point, MHD jets undergo a huge lateral expansion leading to $|B_\phi |>> B_p$ and $m>>1$, which reduces drastically the growth rate of any CD and KH modes, contrary to cylindrical jet configurations (see e.g. \citealt{rose00,moll08,McKi09,port15,kim16}). We thus expect the triggering of CD or KH instabilities only at an altitude $z_I$ where the jet radius has achieved some limiting value (i.e. a quasi-cylindrical jet), determined by the transverse equilibrium with its surrounding medium. 

Our analyzis points to the capital role played by the magnetic surface anchored at $r_J$. That surface corresponds to the last self-confined magnetic surface and thus to a maximum of the electric current flowing inside the jet. But since a sub-FM flow (probably even sub-Alfv\'enic) is expected to be established in the outer magnetic sheath (the grey portion of the magnetic flux anchored beyond $r_J$ in Fig.\ref{fig1}), a velocity shear is also necessarily present. Moreover, since the lateral equilibrium must be fulfilled, the outer (gas plus magnetic) pressure must balance both the inner jet ram pressure and magnetic deconfining force acting on the magnetic sheath. A density increase is thus to be expected in this region. These two elements (inner high speed jet surrounded by a denser and slower outflow) are very similar to those found at the spine-jet interface and clearly observed in GRMHD simulations. In accordance with our discussion, a CD kink instability leading to jet wobbling and wiggles is indeed seen to develop (see e.g. \citealt{McKi09,tche11}). Besides, the instability does not lead to jet destruction, probably because of the existence of this velocity shear \citep{mizu14,kim16,kim18}. 

We argued above that the optimal location for triggering a jet instability leading to jet wobbling would be the altitude $z_I$ where the jet radius would reach its maximal size. This is a situation expected in self-similar Blandford-Payne jets and their generalization, as these type of jets   
always undergo a magnetic recollimation toward the axis \citep{cont94a,ferr97,polk10}. Now, the existence of this intrinsic recollimation may cause a pressure mismatch at the jet-sheath interface (the white zone depicted in Fig.\ref{fig1}), leading to a possible "recollimation" instability \citep{mats13,mats17}. This new 3D instability, first associated to a Rayleigh-Taylor instability, is actually a variant of the centrifugal instability emerging along curved streamlines \citep{gour18a,gour18b}. Whether it could lead to jet destruction and not only jet wobbling as assumed here remains to be fully assessed. However, it has been recently shown that it could be quenched with the presence of some external azimuthal magnetic field \citep{mats21}. 

As a final note, we remark that if we assume the recollimation point to be the locus of the jet instability, then the non-relativistic jet solutions that fulfill all constraints (super-FM, large radius and Eq.(\ref{eq:m})) require a quite tiny mass loading parameter along that surface, namely a disk ejection efficiency $\xi$, defined as $\dot M_{acc} \propto r^\xi$ in the JED, satisfying $5\, 10^{-3} < \xi < 10^{-2}$ (see Figs. 6 and 8 in \citealt{ferr97}). These values are consistent with jet speed estimates done for Cygnus X-1 \citep{petr10}. Note that this range stems from exact mathematical solutions but that more realistic ones should deviate from a strict self-similarity. Indeed, although the JED outermost magnetic surface should fulfill that constraint, the mass loading (or ejection index $\xi$) can increase progressively toward the center.

\section{Discussion}

\subsection{Numerical simulations}

Our wobbling jet scenario is quite generic and should therefore be observed in numerical jet simulations. However, as already pointed out, not only the physical conditions envisioned here are quite demanding but they also require full 3D simulations. Note that this jet wobbling should not be confused with jet precession due to the black hole tilt and leading to a Bardeen-Petterson like (probably modified by large scale magnetic fields) alignment of the innermost disk regions \citep{mcKi13,lisk19,lisk21}. Although such precession may provide the correct physical ground for high frequency (kHz) QPOs, it cannot explain LF QPOs discussed here (see also \citealt{kyla20}). Instead, we rely on a body mode instability in the super-FM regime, involving the whole jet radius and leading to quasi-periodic variations in the jet direction. 

Although the stability of two-component magnetized jets are currently under study (see e.g. \citealt{mill17} and references therein), they usually focus on the interplay between the inner relativistic spine and the outer slower and denser flow (usually referred to as the sheath in these works). We also note that with a much simpler jet profile, kink instabilities can indeed be obtained, affecting the large scale jet morphology and giving rise to interesting quasi-periodic radiative signatures (e.g. \citealt{tche16,barn17,dong20} see also the large scale 2D simulations of \citealt{chat19} showing pinch instabilities). However, we stress that a self-confined magnetized outflow must carry its own electric poloidal current and must therefore be seen as made of three components embedded in an external medium, not just two: the axial spine, the jet and the outer sheath. 

In our view, the role of the central spine in the whole outflow collimation remains to be assessed (see for instance \citealt{barn22}). The jet corresponds here to the super-FM Blandford-Payne outflow launched from the JED. The outer sheath is built up from the unavoidable magnetic flux threading the SAD region, which is settled beyond the JED. Note that such a magnetized sheath must be understood as being part of the whole outflow, since it must carry the return electric current (at the jet-sheath interface) that insulates the jet from the external medium (or cocoon). Understanding the propagation and stability properties of such stratified outflows is of great importance for accreting objects. But addressing both the large spatial and time scales involved, while also accounting for the complex jet stratification imposed by the underlying hybrid disk configuration, remains a numerical challenge.

\subsection{Jet disruption and JED disappearance}

Super-FM jets from JEDs are expected to start wobbling at some altitude $z_I$, but would they be disrupted? This is a difficult question to answer theoretically, as the non-linear saturation could be a simple readjustment of the gradients and the smearing out of the instability. There are for instance circumstances where a CD instability leads to the stabilisation of KH modes \citep{baty02}. Based on the observation that Type-C QPOs are maintained as long as there are steady compact jets (see however discussion in \citealt{fend09, fend14}), we assumed in our simplified picture that this instability is not leading to jet disruption. This requires that the jet is able to propagate up to a distance $L= x r_{jet}$ with, say $x$ of a few hundreds, on a time $T=L/u_{p,max}$ shorter than the characteristic growth time of the instability $\tau = \nu_I^{-1}$ (e.g. \citealt{kim16}). In this expression, $u_{p,max}$ is the maximum poloidal speed of the jet and is thus associated to the ISCO radius $r_{isco}$ (for simplicity, we neglect here the effect of the fast spine). As a consequence, the instability will lead to jet destruction if $\tau < T$, which provides the condition $u_{p,max}/u_{p}(r_j) < x/2n$, where $n= u_p/V_{VM}$ is the FM Mach number measured at the outer jet region. This crude estimate shows right away that as $r_J \rightarrow r_{isco}$ it becomes harder to maintain stability: the thinner the jet and the more fragile, prone to destruction, it gets. This is qualitatively consistent for instance, with the fact that as the jet radius shrinks, the turbulent boundary layer due to a KH surface mode becomes of the order of the jet radius, leading thereby to its disruption \citep{baty06, kim16}. We thus expect a non-destructive jet wobbling whenever $r_J \gg r_{isco}$. 

On the contrary, as the JED shrinks (hard-to-soft transition), the resulting jet would become too narrow to survive the instability. That would be coincident with a dramatic jet restructuration until its complete disruption and disappearance (assumed to occur at the "jet line"). However, before the complete jet disappearance, reconnecting events of the magnetic structure are expected to lead to the dramatic release of jet energy, possibly associated with discrete ejecta and subsequent radio flares \citep{homa20,wood21}. 

These evidences of a link between the appearance of Type-B QPOs and the launch of discrete ejecta fits quite well within our jet instability scenario for QPOs. As $r_J \rightarrow r_{isco}$, we expect the jet to become also narrower due to the (relatively stronger) external magnetic pressure \citep{spru97}. This translates into a QPO frequency $\nu_I \sim \Omega_K(r_J)/(2m)$ (Eq.~\ref{eq:nu_i}) that could change while the transition radius $r_J$ remains almost constant. Indeed, a smaller opening of the magnetic field lines leads to a smaller jet acceleration, hence to a smaller Alfv\'enic Mach number $m$. As a consequence, the QPO frequency may (slightly) increase while there is no detectable change in the inner cold disk radius \citep{kara19,ma21}. This is an aspect of jet collimation that has not been fully investigated yet.

\subsection{Detectable signatures: IR QPOs, jet wobbling}  

Within the framework of wobbling jets, getting QPOs at the same frequency and at energy bands usually associated with jet emission (ie OIR) appears rather natural. Indeed, the jet instability is triggered at an altitude $z_I$ (the recollimation zone) and gives rise to perturbations that propagate both up-  and downstream. It will take a time $\Delta t_X$ for the upstream waves to travel down to the disk and give rise to an observable QPO in the X-ray band. But since jet wobbling is quasi-periodically produced, it may well give rise to a OIR QPO at a distance $z_{IR} > z_I$ after a time $\Delta t_{IR}\sim (z_{IR}-z_I)/u_p$, where $u_p$ is the flow speed (and possibly to some variability signature in radio bands \citealt{teta19}). There is therefore no obvious reason why $\Delta t_X$ should always be shorter than $\Delta t_{IR}$, so we do not expect any clear trend for the time lag between X and OIR QPOs (see also \citealt{vele13, vele15}).  
 
This is consistent with OIR QPOs lagging behind X-ray QPOs with $\sim 0.1$ sec in several XrBs \citep{gand10,kala16,gand17}, but also with the non detection of any lag in MAXI J1535-571 \citep{vinc21} or even X-rays possibly lagging behind the optical QPO in MAXI J1820+070 \citep{paic21}. Moreover, there are evidences of different properties between the two QPOs, like the rms-flux relation \citep{vinc18} or the time evolution of the power spectral density \citep{vinc19}. Although a common origin for X and OIR QPOs is quite natural within our framework, it seems plausible that some filtering effect is occurring as perturbations are propagating downstream the jet (see also \citealt{hyne03}).

Probing jet wobbling in BH XrB jets may be quite hard to achieve because of the lack of resolution (see however \citealt{mill19}). In the context of AGN, this requires instead a long term monitoring of jets. It is well known that AGN jets display non-radial motions, indicative of accelerated, non-ballistic motions (see e.g. \citealt{list16, bocc17b}). But jet wobbling has been inferred in various BL Lac sources implying time scales from 2 to 20 yr (e.g. \citealt{agud12b, arsh20} and references therein). \citet{walk18} have also found evidences of such a pattern in the radio galaxy M87, with a period of $\sim 9$ yr. When scaled down to a 10 solar mass BH, this value provides a frequency around 2 Hz, which is consistent with the range of LF QPOs in BH XrBs. 

In the realm of Young Stellar Objects, the situation is quite similar. Protostellar jets seen in the optical are thought to arise from the innermost disk regions, where orbital periods are ranging from few days to a year or so (\citealt{ferr06b,ray07} and references therein). On the other hand, these jets harbor knots that resemble bow shocks whose separations are indicative of time scales ranging from years to tens of years (e.g.\citealt{lope03,agra11,elle14,lee17,tabo17}). Such time scales are too short to be explained by perturbations due to a companion star, and too long to be related to any dynamical time scale at the launching radius (see for instance jet wobbling in HH30, \citealt{louv18} and references therein). Instead of relating these knots to time variable ejection events from the source, we propose to relate them to the same jet instability. Such instability could be very similar to the one invoked here to explain LF QPOs around compact objects, although leading possibly to jet disruption with a subsequent ballistic motion in some sources.

\section{Conclusion}

Building upon the success of the JED-SAD framework to reproduce the radio emission and X-rays spectral energy distributions during various cycles of the archetypal source GX 339-4, we addressed in this paper the question of the origin of Type-C LFQPOs. 

We first critically analyzed the most common models invoked for such QPOs: 1) models looking for a specific process triggered at the transition radius $r_J$, 2) the accretion-ejection instability and 3) the solid-body Lense-Thirring disk precession model. We showed that the first two types of models are facing major theoretical issues, unsolved yet, and that the published versions of these models do not account for the observed tight correlation with the disk transition radius (Eq.~\ref{eq:rJ}). We also argued that these models could not be operating within the JED-SAD framework. 

We then discussed the case of the solid-body LT-precession model, which is invoked whenever the black hole spin is misaligned with the disk angular momentum vector. We argued that evidences in GRMHD simulations for such a solid-body behavior were not fully convincing, especially in the case of highly magnetized accretion flows. This casts thereby doubts on the capability of this model to operate within the JED-SAD framework (at least for moderate black hole tilts). As suggested however by current numerical simulations, the innermost JED region, up to a few $r_g$, would probably be aligned with the black hole spin, leading to an inner jet precession (namely a precession of both the Blandford-Znajek axial spine and the innermost zones of the surrounding Blandford-Payne jet). But whether some fraction of the JED, located beyond the aligned region, would undergo a solid-body LT-precession remains an open issue that deserves further numerical simulations,  long enough and tailored to maintain an outer cold thin disk. We argued however that such an effect would hardly be efficient up to the transition radius $r_J$ with the outer SAD, as required by observations.    

We propose instead that the JED-SAD framework offers the conditions allowing for a jet wobbling, namely a non-destructive long wavelength 3D body mode probably triggered by a kink or recollimation instability in the super-FM jet regime. Such a wobbling, triggered away from the disk, will nevertheless affect the disk through FM waves travelling upstream along the surrounding magnetic sheath, as well as, of course, downstream the jet. This outer magnetic sheath is the key ingredient allowing to connect the fast inner jet, which is hammering quasi-periodically against the sheath, to the underlying resonant disk (JED). This scenario offers thereby a unique explanation to the existence of QPOs seen both in disk (X-rays) and jet (UV and OIR) emission signatures. 

The theoretical foundations of this scenario are threefold: i) existence of exact calculations of MHD accretion-ejection flows with the correct required properties; ii) self-consistent thermal disk balance calculations, with thorough confrontations to observations; iii) current knowledge on MHD instabilities in stratified super-FM flows. Notwithstanding these facts and hints, the proposed dynamical mechanism as well as its radiative consequences remain to be firmly established. To do so, global 3D jet simulations must be designed so that such an instability could be indeed observed, with its up-streaming perturbations heading toward the disk. On the radiative side, the work of \citet{caba10} should be extended in order to incorporate the existence of these incoming perturbations, and possibly address also the question of the time lags and their energy dependence.       
 
We finally note that polarization measurements by IXPE, just launched in Dec. 2021 or, if selected, and in a more distant future, by eXTP, may bring great enlightenment on LFQPOs. Indeed, if swings in polarization angle are detected, this will provide stringent conditions that all the models and scenarios discussed above will have to explain. This is provided of course that such constraints are usable, as required integration times are much bigger than the periods to probe (but see for instance \citealt{ingr17b}).

\begin{acknowledgements}
JF would like to thank Adam Ingram and Matthew Liska, as well as the referee, Chris Done, for interesting discussions that allowed to improve the quality of the paper. We acknowledge financial support from the CNES French space agency and PNHE program of French CNRS.
\end{acknowledgements}

\bibliographystyle{aa}

\end{document}